\documentclass[aps,prl,twocolumn,showpacs,preprintnumbers]{revtex4}

\usepackage{graphicx}

\let\bm=\bibitem

\def\ft#1#2{{\textstyle{{\scriptstyle #1}\over {\scriptstyle #2}}}}
\def\fft#1#2{{#1 \over #2}}
\def\by{{\bf y}}
\def\bx{{\bf x}}
\def\del{\partial}
\def\nn{\nonumber}
\def\sst#1{{\scriptscriptstyle #1}}
\def\0{{\sst{(0)}}}
\def\1{{\sst{(1)}}}
\def\2{{\sst{(2)}}}
\def\3{{\sst{(3)}}}
\def\4{{\sst{(4)}}}
\def\5{{\sst{(5)}}}
\def\6{{\sst{(6)}}}
\def\7{{\sst{(7)}}}
\def\8{{\sst{(8)}}}

\newcommand{\bea}{\begin{eqnarray}}
\newcommand{\eea}{\end{eqnarray}}
\newcommand{\be}{\begin{equation}}
\newcommand{\ee}{\end{equation}}

\begin{document}

\preprint{DAMTP-2005-3\ \ \ \ MIFP-05-01\ \ \ \ {\bf hep-th/0501117}}

\title{Brane Worlds in Collision}

\author{G.W. Gibbons,$^1$ H. L\"u,$^{2}$ C.N. Pope$^{2}$}
\altaffiliation[]{Research is supported in part
by DOE grant DE-FG03-95ER40917}
\affiliation{%
${}^1\!\!\!$ DAMTP, Centre for Mathematical Sciences, Cambridge University
Wilberforce Road, Cambridge CB3 OWA, UK\\
${}^2\!\!\!$ George P. \& Cynthia W.
Mitchell Institute for Fundamental Physics,
Texas A\&M University, College Station, TX 77843, USA
}%

\date{January 14, 2005}

\begin{abstract}

     We obtain an exact solution of the supergravity equations of
motion in which the four-dimensional observed universe is one of a
number of colliding D3-branes in a Calabi-Yau background.  The
collision results in the ten-dimensional spacetime splitting into
disconnected regions, bounded by curvature singularities.  However,
near the D3-branes the metric remains static during and after the
collision.  We also obtain a general class of solutions representing
$p$-brane collisions in arbitrary dimensions, including one in which 
the universe ends with the mutual annihilation of a positive-tension 
and negative-tension 3-brane.

\end{abstract}

\pacs{11.25.-w, 98.80.Jk, 04.50.+h}
\maketitle

   The idea that our spacetime is a brane in a higher-dimensional
world leads naturally to the suggestion that the big bang is the
result of a brane collision \cite{dvatye,khovsttu}.  Colliding
brane-world scenarios offer an alternative to conventional
four-dimensional cosmological models, and an important challenge is to
make predictions that distinguish between intrinsically
higher-dimensional effects and conventional lower-dimensional
physics. In order to do so, a better understanding of brane-brane
interactions is needed.  Studies of D3-D3
and D3-{$\overline{\hbox{D}}$}3 collisions have so far been restricted to
approximations based on effective field theories on the brane, or on
perturbative string-theory techniques \cite{tupest,mcamit}.  There
have been no exact treatments based on the supergravity equations of
motion. The purpose of this letter is to provide an exact
treatment in the case of colliding D3-branes, moving in a Calabi-Yau
background.  Our technique extends to more general brane-collision
processes, which we shall also discuss briefly.  It requires an
extension to $p$-branes of an analysis of black-hole collisions
developed in \cite{kastra,brhokatr}, which made use of a remarkable
time-dependent generalisation \cite{kastra} of the static
Majumdar-Papapetrou metrics \cite{majpap}.  The collision of extreme
Reissner-Nordstr\"om black holes had been discussed previously using a
modulus-space description \cite{girub,feea}.  The importance of the
work of \cite{brhokatr} was that at the expense of introducing a
cosmological term, the dynamics could be studied {\it exactly}.

     The Kastor-Traschen solution \cite{kastra} of Einstein-Maxwell
theory with cosmological constant $\Lambda>0$ takes the form
\bea
ds_4^2 &=& -H^{-2}\,dt^2 + H^2\, d\by^2\,,\qquad
A = H^{-1}\, dt\,,\nn\\
H&=& h\, t + \Phi(\by)\,,\qquad \Phi(\by)\equiv
\sum_k \fft{M_k}{|\by -\by_k|}\,,
\eea
where $h=-\sqrt{\Lambda/3}$.  Note that if the masses $M_k$ are
set to zero, the solution
is de Sitter spacetime, and the apparent singularity at $t=0$ is a
mere coordinate artefact.  (It corresponds to a null hypersurface
separating the maximally-extended de Sitter spacetime into two
antipodally-related parts, with $t>0$ and $t<0$.
In fact, the involution $t\rightarrow -t$ is the antipodal map.)  

   When $t$ is negative, the surfaces $t=$ constant are everywhere
spacelike.  Near each $\by_k$, there is an infinite throat which is
very similar to the throat of the asymptotically-flat extremal
Reissner-Nordstr\"om solution, for which $H=1+\Phi(\by)$.  Far from
the throats, however, there is a contracting $k=0$ de Sitter universe
$ds_4^2 \sim -d\tau^2 + a^2(\tau)\, d\by^2$, 
with $a(\tau)= e^{-h\tau}=ht$, and $\tau$ going to $+\infty$.

    When $t$ is positive, however, one cannot go too far away from the
throats before the function $H$ becomes negative.  In fact, the
time-dependent locus $H=0$ is a spacetime singularity.  In
\cite{brhokatr}, a detailed discussion is given of the locations of
the horizons that form around the black holes.  The simplest case to
consider would be that of two black holes, of masses $M_1$ and $M_2$,
with $M=M_1+M_2 < \ft14 |h|^{-1}$.  This represents the head-on
collision of two black holes.  As $t$ approaches zero, one has two
outer apparent horizons which ultimately coalesce, and a single
time-independent Reissner-Nordstr\"om-de Sitter black hole is formed
at late times \cite{brhokatr}.

     After the appearance of \cite{kastra}, Maki and Shiraishi
obtained a general class of time-dependent multi-particle charged
solutions \cite{makshi}, coupled to a scalar field $\phi$.  In
general, the scalar field couples to the cosmological constant in
Liouville-like fashion. Some four-dimensional examples were analysed
in \cite{horhor}.  In some cases the cosmological constant can be set
to zero. The modulus-space metrics for these multi-particle solutions
were given in \cite{shir}.  The slow motion of NS-NS 5-branes was
discussed in \cite{felsam}.  In \cite{fregibsch}, a particular case
was used to study an expanding gas of D0-branes, allowing a comparison
of the modulus-space approximation with the exact solution of
\cite{makshi}.

    In order to discuss brane scattering, we shall lift some of the
Maki-Shiraishi solutions to higher dimensions.  Their Lagrangian
in $(N+1)$ dimensions is
\be
{\cal L} = R -\fft{4}{N-1}\, (\del\phi)^2 - e^{-4a\phi/(N-1)}\, F^2
  - \Lambda\, e^{4b\phi/(N-1)}\,.
\ee
We restrict attention to solutions described by Case (II) in
\cite{makshi}, for which
\bea
&&ds^2 = - V^{-\ft{2(N-2)}{N-2+a^2}}\, dt^2 +
     a^2(t)\, V^{\ft{2}{N-2+a^2}}\, d\by^2\,,\nn\\
&&A = -\sqrt{\fft{N-1}{2(N-2+a^2)}}\, \fft{1}{\sqrt{C}\,
  (a(t))^{a^2}\,} V^{-1}\, dt\,,\nn\\
&&e^{-4a\phi/(N-1)} = C\, (a(t))^{2a^2}\, V^{2a^2/(N-2+a^2)}\,,
\eea
where
\bea
V &=& 1 + \sum_k \fft{\mu_k}{\sqrt{C}\, (a(t))^{N-2+a^2}\, (N-2)\,
   |\by - \by_k|^{N-2}}\,,\nn\\
a(t) &=& \Big(\fft{t}{t_0}\Big)^{1/a^2}\,,\quad
\fft{N}{a^2} - 1 = \fft{a^2\, t_0^2\, \Lambda}{C\, (N-1)}\,.\label{cosrel}
\eea
 
   By contrast with the Kastor-Traschen case, the presence of the
scalar field allows us to take $\Lambda=0$, by choosing $a^2 = N$.  In
the notation of \cite{luposest}, this coupling strength of the dilaton
to the Maxwell field is characterised by the constant $\Delta$ taking
the value 4.  In fact $\Delta=4$ is precisely the coupling that one
encounters, in any dimension, for a Maxwell field arising from the
Kaluza-Klein vector in a circle reduction \cite{luposest}.  It is also
the coupling strength for {\it any} field strength in maximal
supergravity in any dimension \cite{luposest,dukhlu}.  Thus, for
example, if $N=9$ the vector field is the Ramond-Ramond vector of type
IIA string theory, which may also be interpreted as arising from
M-theory via Kaluza-Klein reduction \cite{fregibsch}.

    To construct colliding D3-branes we require $N=6$, and the Maxwell
field strength may be viewed as arising from the 5-form of type IIB
supergravity.  Lifting the solution from seven to ten dimensions, we
obtain a time-dependent D3-brane solution, in which the only
non-trivial fields are the metric and the 5-form, given by
\bea
ds_{10}^2 &=& H^{-1/2}\, (-dt^2 + d\bx^2) + H^{1/2}\,
      d\by^2
\,,\nn\\
F_\5 &=& dt\wedge d^3\bx\wedge dH^{-1} +
             *(dt\wedge d^3\bx\wedge dH^{-1})\,,\nn\\
H&=& h\, t + \Phi(\by)\,,\qquad \Phi(\by)\equiv
\sum_k \fft{M_k}{|\by -\by_k|^4}\,,\label{d3met}
\eea
where $h$ is an arbitrary constant (which we shall take to be negative,
to parallel the discussion in \cite{brhokatr}). Sending $t\rightarrow t+1/h$
one gets a standard supersymmetric static D3-brane if $h=0$.  However, the
time-dependent $h\ne0$ solution is non-supersymmetric. 

    If the non-negative constants $M_k$, which correspond to the
3-brane charges, all vanish, then we obtain a vacuum solution of the
Einstein equations, which is a generalisation of the familiar Kasner
solution,
\bea
ds_{10}^2 &=&  (ht)^{-1/2}\, (-dt^2 +  d\bx^2) + (ht)^{1/2}\,
       d\by^2 \,,\nn\\
&=& - d\tau^2 + \Big(\fft{3h\tau}{4}\Big)^{-2/3}\, d\bx^2 +
         \Big(\fft{3h\tau}{4}\Big)^{2/3}\, d\by^2\,,
\eea
where we have defined $h\tau = \ft43 (ht)^{3/4}$.  
The singularity at $t=0$
is, by contrast with the de Sitter case, a true curvature singularity,
and the Kasner metric cannot be continued from negative to positive
$t$.  In the ten-dimensional metric the three spatial dimensions are
increasing in size as $t$ increases towards zero from an initial
negative value.  From the point of view of four dimensions, the
interpretation is simplest in the Einstein conformal gauge, which is
achieved by writing
\be
ds_{10}^2 = e^{2\alpha\varphi}\, ds_4^2 + e^{-2\alpha\varphi/3}\,
     d\by^2\,,
\ee
where $\alpha^2 = 3/16$.  Introducing the Einstein proper-time
coordinate $T= 3h\tau^2/8$, the four-dimensional
Einstein-frame metric becomes
\be
ds_4^2 = -dT^2 + (\ft32 h T)^{2/3}\, d\bx^2\,.\label{frle}
\ee
This is of the form expected for gravity coupled to a massless scalar
field, which behaves like a perfect fluid with a stiff-matter equation
of state for which the energy density equals the pressure.  In the
Einstein conformal gauge, the spatial 3-sections {\it contract} as $T$
increases towards zero from an initial negative value; $T=0$ is a
big-crunch singularity.  If instead we run time backwards, the
solution represents an expanding universe with a big-bang singularity.
The discussion above emphasises the point that four-dimensional
physics should be analysed in the Einstein conformal frame.
                                                                               
   Now let us consider the case where the constants $M_k$ are
non-zero.  Starting with $t$ negative but increasing towards zero, the
solution represents 3-branes moving in a background Kasner universe,
in which, measured in the ten-dimensional metric, the transverse space
contracts while the D3-brane world volume expands.  From the point of
view of four dimensions, the situation is similar to the pure Kasner
case discussed above, as long as one stays well away from all of the
D3-branes, because the relevant conformal factor is
\be
e^{-2\alpha\varphi} = (ht)^{3/2} \, \Big(1 + \fft1{ht}\,
   \sum_k \fft{M_k}{|\by - \by_k|^4}\Big)^{3/2}\,.
\ee
At any fixed negative $t$ the surfaces of constant time are smooth,
asymptotically flat and non-compact.  Near each 3-brane,
$\by\sim\by_k$, the 10-metric is static, and well-approximated by the
product metric on AdS$_5\times S^5$. Because the transverse directions
are inhomogeneous, the four-dimensional interpretation depends upon
where one is in the transverse space.
 
   The situation as $t$ goes to zero is more complicated than in the
pure Kasner case.  If $t$ is exactly zero, we obtain a non-singular
ten-dimensional configuration that is not asymptotically flat, which
may be thought of as a union of AdS$_5\times S^5$ components.  An
analogous solution has been discussed for AdS$_2\times S^2$ in
\cite{brill,strmicmal,ngpe}, in connection with anti-de Sitter
fragmentation.
                                                                               
   If the charges $M_k$ are non-zero the solution continues to exist
in a neighbourhood of each 3-brane when $t$ becomes positive, by
contrast with the Kasner solution.  Specifically, the metric is
\bea
ds_{10}^2 &=& (-ht)^{-1/2}\, \Big(\fft{\Phi(\by)}{(-ht)} - 1\Big)^{-1/2}\,
   (-dt^2 + d\bx^2)\nn\\
&& +(-ht)^{1/2}\, \Big(\fft{\Phi(\by)}{(-ht)} - 1\Big)^{1/2}\, d\by^2\,,
\eea
where $\Phi(\by)$ is given in (\ref{d3met}).  For all positive times,
the metric exists inside a domain $D_t$ bounded by the level set
$\Phi(\by)=-ht$.  Inside the domain $D_t$, but away from the D3-branes
themselves, the transverse dimensions expand while the three spatial
dimensions contract even with respect to the four-dimensional
Einstein-frame metric, if one stays at fixed $\by$. However, if one
moves in the transverse space in such a way that $\Phi(\by)/(-ht)$
remains approximately constant (and greater than 1), then the
corresponding four-dimensional universe has expanding spatial
sections.

\begin{figure}
\includegraphics{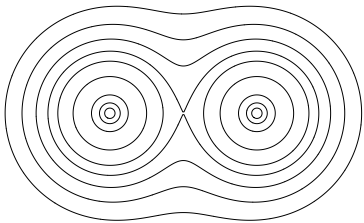}
\caption{\label{figure1} The level sets of $\Phi(\by)$ for two mass
  points: The universe after
the collision is confined inside a level set $\Phi(\by)>-ht$.  As time
progresses, the level set $\del D_t$ shrinks and splits into
two components, which then shrink around the two D3-branes.}
\end{figure}

    At small positive $t$, the domain $D_t$ is a large connected ball
containing all the D3-branes.  As $t$ increases, the domain $D_t$
shrinks, and eventually splits into disconnected pieces which shrink
down onto each mass point.  In other words, the universe splits into
disconnected regions containing the individual D3-branes.  Near each
D3-brane, however, the metric is static, and again well approximated
by the product metric on AdS$_5\times S^5$.

    Consider, for example, the case of two mass points, which
is illustrated in Fig.~\ref{figure1}.  At negative
times the D3-branes are approaching one another in a head-on
collision, with the universe contracting about them.  The D3-branes
never actually meet, but rather scatter off each other into a universe
bounded by spacetime singularities.

    In the solutions we have discussed so far, the six-dimensional
transverse space is flat.  In more realistic examples one might want
to consider D3-branes moving in a Calabi-Yau manifold.  In fact, the
solutions given by (\ref{d3met}) may be generalised to the case where
the flat transverse metric $d\by^2$ is replaced by any Ricci-flat
6-metric, and $H=ht + \Phi(\by)$ where $\Phi(\by)$ is any harmonic
function on the Ricci-flat 6-manifold.  In particular, we can take the
Ricci-flat manifold to be Calabi-Yau.

   The time-dependent D3-brane solutions that we obtained above admit a
straightforward generalisation to the case of an arbitrary $p$-brane
supported by a $(p+2)$-form field strength with a dilaton coupling
characterised by $\Delta=4$ in the notation of \cite{luposest}.  The
relevant $D$-dimensional Lagrangian can be expressed as
\be
{\cal L} = R -\ft12 (\del\phi)^2  - \ft1{2 (p+2)!}\,
   e^{c\phi}\, F_{(p+2)}^2\,,
\ee
where
\be
c^2 \equiv \Delta - \fft{2(p+1)(D-p-3)}{D-2} = 4 -
\fft{2(p+1)(D-p-3)}{D-2}\,.
\ee
The time-dependent $p$-brane solution is given by
\bea
\!\!\!\!\!\!&&ds^2 = H^{-\ft{D-p-3}{D-2}}\, (-dt^2\!\! +\!\! d\bx^2) +
    H^{\ft{p+1}{D-2}}\, g_{mn}(y)\, dy^m\, dy^n\,,\nn\\
\!\!\!\!\!\!&&F_{(p+2)} = dt\wedge d^p\bx \wedge dH^{-1}\,,\nn\\
\!\!\!\!\!\!&& \phi = \ft12 c\, \log H\,,\quad H = ht + \Phi(y)
\eea
and $h$ is an arbitrary constant.  Here $g_{mn}(y) \, dy^m\, dy^n$ is
any Ricci-flat $(D-p-1)$-metric, and $\Phi(y)$ is an harmonic function
in this metric.  One might, for example, take the flat
metric $g_{mn}(y) \, dy^m\, dy^n= d\by^2$ and $\Phi(\by) =
 \sum_k M_k/|\by - \by_k|^{D-p-3}$.

   These solutions can be derived by lifting the Case (II)
time-dependent Maki-Shiraishi solutions \cite{makshi}, setting $a^2=N$
so that $\Delta=4$.  Alternatively, they can be derived by using
T-duality symmetries to relate them to pp-waves.  We begin by noting
that the pp-wave metric
\be
ds^2 = 2 dx^+\, dx^- + \fft{4}{(2h x^+)^2}\, \Phi(y)\, (dx^+)^2
    + g_{mn}(y)\, dy^m\, dy^n
\ee
is Ricci-flat for any $\Phi(y)$ that is harmonic in the Ricci-flat
metric $g_{mn}(y)\, dy^m\, dy^n$.  It is invariant under the boost
given by $x^+\rightarrow \lambda\, x^+$, $x^-\rightarrow
\lambda^{-1}\, x^-$.  We may therefore perform a Kaluza-Klein
reduction on this boost symmetry by first introducing coordinates
$(t,z)$, given by $x^- = t\, e^{\ft12 h z}$ and 
$x^+ = -(2/h) e^{-\ft12 h z}$. This gives
\be
ds^2 = -H^{-1} dt^2 + H (dz + H^{-1}\, dt)^2 + g_{mn}(y)\, dy^m dy^n\,,
\ee
with $H= h\, t + \Phi(y)$.  Reducing on the boost Killing vector
$\del/\del z$ gives a time-dependent 0-brane.  If one is starting from
$D=11$ supergravity, this gives the D0-brane that was discussed in
\cite{fregibsch}, which is the special case $N=9=a^2$ of Maki and
Shiraishi.

   By performing a sequence of T-duality transformations in the
standard way (see, for example, \cite{bergs}), this immediately gives
us all the time-dependent $p$-brane solutions that we have presented
above.  A further extension is to consider 
harmonic intersecting $p$-branes.  It is straightforward to see that
any one (but only one) of the harmonic functions can be generalised
from $1+\Phi(y)$ to $ht+\Phi(y)$.

   A further generalisation is to apply a discrete electric/magnetic
duality transformation in four dimensions.  Consider, for example, the
metric representing a maximally electrically charged Kaluza-Klein
black hole:
\bea
ds_4^2 &=& - H^{-1/2}\, dt^2 + H^{1/2}\, d\by^2\,,\nn\\
F &=& -dt\wedge dH^{-1}\,,\quad \phi= \ft12 \sqrt3\, \log H\,,
\label{4meta}
\eea
with $H= ht + \Phi(\by)$.  After making the duality transformation we
obtain the magnetic solution, with the same metric and with
\be
F= dA={*_3 d\Phi}\,,\qquad \phi = - \ft12 \sqrt3\, \log H
\ee
where $*_3$ denotes the Hodge dual in the flat 3-metric $d\by^2$.
Lifting to five dimensions, we obtain the Ricci-flat metric
\be
ds_5^2 = - dt^2 + H\, d\by^2 + H^{-1}\, (dz+ A)^2\,.
\label{5met}
\ee
The last two terms in (\ref{5met}) give a time-dependent sequence of
hyper-K\"ahler 4-metrics.  The interpretation of the metric
(\ref{4meta}) is that it represents an assembly of
freely-expanding point particles in a background Friedman universe
with a scale factor $a(T)$ proportional to $T^{1/3}$.  In the 
slow-motion approximation the motion is free and hence the branes separate 
linearly with time \cite{ruback}.

   It would be of great interest to study the stability of Ho\v rava-Witten
brane models by finding exact time-dependent solutions; some preliminary
results are reported in \cite{ccglp}.  Here, we present a different colliding 
brane-world model, which however captures the essence of all brane-world
instabilities. We take $\Phi(\by)=M |y_1|$ in (\ref{d3met}), and reduce on
$(y_2,\ldots,y_6)$, obtaining the 5-metric
\be
ds_5^2 = -(ht + M |y_1|)^{1/3}\, (-dt^2 + d\bx^2) + 
(ht+ M |y_1|)^{4/3}\, dy_1^2\,,
\ee
which describes a negative-tension time-dependent 3-brane at $y_1=0$,
supported by a 5-form. If the solution is reflected about $y_1=L$, and
if $y_1=2L$ is identified with $y_1=0$, we obtain an $S^1/Z_2$
orientifold with an additional positive-tension brane at $y_1=L$.  The
proper distance between the 3-branes is $6[(ht+ML)^{5/3}-
(ht)^{5/3})]/(5M)$.  This decreases monotonically as $t$ increases to
0.  Defining $T=2(ht+ML)^{3/2}/(3h)$, the Einstein metric induced on
the positive-tension brane is of the Friedman-Lemaitre form
(\ref{frle}), driven by a massless scalar field, {\it i.e.} 
the radion.  At $t=0$ a singularity forms on the
negative-tension brane, and moves towards the positive-tension brane,
causing the complete annihilation of the universe at $t= ML/(-h)$,
i.e. $T=0$. This is the way the brane-world ends, not with a whimper but a
bang.


\begin{acknowledgments}

   G.W.G. is grateful to Indrajit Mitra for helpful discussions, and to
the George P. \& Cynthia W. Mitchell Institute for Fundamental Physics
for hospitality during the course of this work.

\end{acknowledgments}

\end{document}